\documentclass{anstrans}
\title{Radiation transport calculations for the European Spallation Source accelerator environment}
\author{Douglas D. Di Julio$^{a}$, Mamad Eshraqi$^{a}$, Wolfgang Hees$^{a}$, Yvonne Hinrichsen$^{a}$, Esben Klinkby$^{a,b}$,
  Anton Lundmark$^{a}$, Gunter Muhrer$^{a}$, and Daniel Noll$^{a}$}

\institute{
  $^{a}$European Spallation Source ERIC, P.O Box 176, SE-221 00 Lund, Sweden, douglas.dijulio@ess.eu, mohammad.eshraqi@ess.eu,
  wolfgang.hees@ess.eu, yvonne.hinrichsen@ess.eu, anton.lundmark@ess.eu, gunter.muhrer@ess.eu,daniel.noll@ess.eu \\
  $^{b}$DTU Physics, Technical University of Denmark, 4000, Roskilde, Denmark, esbe@dtu.dk
}

\usepackage{graphicx} 
\usepackage{booktabs} 
\usepackage{microtype} 

\begin{document}
\section{Introduction}
A central component of the European Spallation Source (ESS) \cite{Garoby_2017} is the high-power proton accelerator. The accelerator aims in the future to provide 2 GeV protons to a rotating tungsten target for the production of neutrons at 5 MW of average beam power for neutron scattering studies. Extensive shielding surrounds the accelerator, in order to provide sufficient safety for the public and workers against radiation produced along it's length. This largely comprises several meters of soil, called the berm, and concrete structures located around the accelerator tunnel. However, due to the need for access to the accelerator, during maintenance and connections for other utilities, the shielding contains a number of penetrations which leads to weaknesses in localized areas. For these reasons, shielding design of such a facility must take care to address issues related to both the deep-penetration of the radiation through thick shields, while at the same time accounting for the streaming nature of the radiation through ducts and chicanes. In addition, radiation produced from activated components of the accelerator also poses risks for workers and public. For example, activated magnets will need maintenance and workers may need to access areas where activated water circulates. Thus, activation analyses are also needed in combination with the prompt-dose rate estimates for a complete analysis.  \\
\indent In order to address the above mentioned issues, a detailed model of the ESS accelerator has been developed and used as the basis for both
prompt-shielding and activation estimates at the ESS. In this summary, we present the results of a series of studies carried out on various components around the ESS accelerator. These studies have been addressed through the use of recent software developments, which include for example the Monte Carlo Particle Lists (MCPL) \cite{kittelmann2017} tool, ADVANTG \cite{Mosher2015}, and the geometry builder CombLayer \cite{Ansell}, for producing highly parametric models and providing a simple interface for transferring geometries between different Monte-Carlo software. Some example applications will be presented below.\\
\section{Methodology}
\subsection{Software Packages Used}
The Monte-Carlo calculations were carried out with either PHITS 3.1 \cite{niita2006} or MCNP 6.2 \cite{mcnp6}. The accelerator source term model was developed
for PHITS 3.1, which also makes it possible to provide activation calculations through the interface provided with the distributed software to DCHAIN-SP \cite{dchain}.
On the other hand, PHITS 3.1 cannot be directly used with ADVANTG for efficient weight-window generation, and thus MCNP6.2 was used where such weight-windows
were needed. \\
\indent There are several inherent challenges with the usage of two Monte-Carlo codes. The first is related to the fact that identical geometries in both simulation packages are needed. This however is made possible through the use of CombLayer. CombLayer is a C++ software toolkit, for which one defines the geometrical model using object-oriented programming techniques. The output of the software is an identical geometry input file for either MCNP 6.2 or PHITS 3.1. \\
\indent The second inherent challenge is how to transfer information from the source term model, computed using PHITS 3.1, to an MCNP6.2 model. This can be accomplished with the MCPL toolkit, which provides a common file format for the transfer of Monte-Carlo particle information from one code to another. A PHITS interface was not available in the original release of MCPL, however one was developed for the purposes of the work described here. \\
\indent Two other software packages are used where needed. The first is ADVANTG for the generation of variance reduction parameters, and a detailed overview of the application of this software at ESS can be found in Ref. \cite{miller2020}. The second is ROOT \cite{root}, which has been used to help construct intermediate source terms. 
\begin{figure*}
  \centering
  \resizebox{0.6\textwidth}{!}{\includegraphics{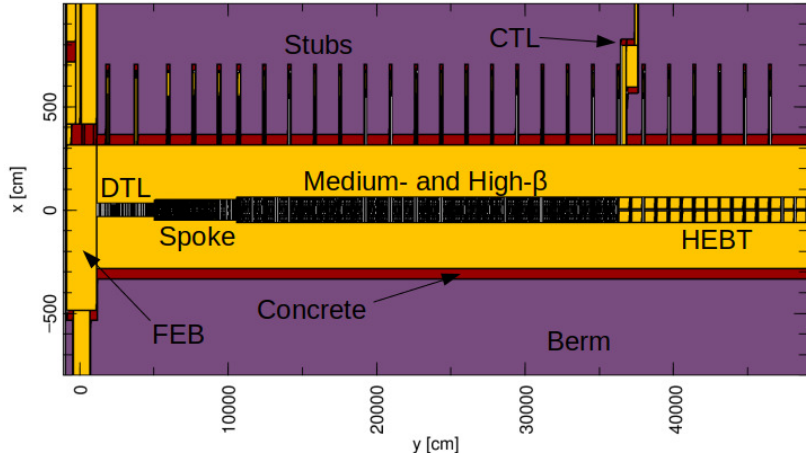}}
  \caption{\label{fig:2} A general overview of the main components around the accelerator tunnel in CombLayer model.}
\end{figure*}
\subsection{Input Data for the Simulations}
The basis for the models used in our studies is the previous work on the ESS accelerator shielding, described in more detail in Ref. \cite{mokhov2016}. These were developed within the framework of the MARS15 \cite{mars1,mars2} code system and the normal-conducting linac (NCL) part was originally ported to CombLayer \cite{batkov2017}. This model was updated to include the higher-energy sections of the accelerator and now contains the NCL, the double-spoke, medium, high-$\beta$ elliptical cavity sections, and the high-energy beam transfer section. The geometries include the individual DTL magnets, spoke cavities, elliptical cavities, and quadruple magnets, with the positions based on lattice files from the accelerator group at the ESS. Generally, these components were taken from the MARS15 model, with updates to the geometries based on the latest CAD data. Surrounding the linac is a tunnel of standard concrete and berm soil. Additionally, the model contains the front-end building (FEB), the geometry for 27 chicanes connecting the tunnel to the Klystron gallery, called the stubs, a cryogenic transfer line (CTL) gallery, and four alignment penetrations. The major components are highlighted in Fig. 1. \\
\indent The source term described in Ref. 15 \cite{mokhov2016} has been adapted as a user-defined source and compiled together with PHITS 3.1. The source term includes the energy dependence of the proton source along the linac in addition to an average beam loss of 1 W/m along it's length. The 1 W/m value is a generally used limit for the average uncontrolled losses within the accelerator community \cite{beamScraping}, in order to allow for hands on maintenance. This limit was conservatively adopted at the ESS and to be applied along the entire length of the accelerator \cite{Garoby_2017}. \\
\indent In the source term implementation, the protons are generated uniformly across the accelerator, according to 1 W/m limit, along a cylinder of radius, defined by the smallest apertures of the magnets in the various sections, and with a low-grazing angle of 3 mrad. While the initial starting points in our model are uniform across the length of the accelerator, in reality due to $\sim$mm thickness of the beam-pipes, a large fraction of the lost protons are incident on the quadruple magnets, resulting in more point-like losses along the accelerator length. \\
\indent For the calculations presented here we have used the default nuclear models in the Monte-Carlo software. This includes in PHITS 3.1 the Intra-Nuclear Cascade of Li\`ege \cite{boudard2013} and the KUROTAMA model \cite{Iida2007}, which provides reaction cross-sections for nucleus-nucleus and nucleon-nucleus reactions. The default nuclear cascade model in MCNP6.2 is the Cascade-Exciton model (CEM) \cite{Gudima1983,Mashnik2008,Mashnik2012}. Generally, proton and neutron libraries were taken from ENDF/B-VII.0/VII.1/VIII \cite{endf7,endf71,endf8} but in some limited cases we supplemented the above libraries with data from TENDL-2017/2019 \cite{tendl,tendl2}, JENDL-4.0 \cite{shibata2011} and JEFF 3.2 \cite{jeff32}. 
\section{Radiation transport calculations}
In the following sections we present results of examples of calculations based on the models shown above. The first is a calculation
of an accelerator penetration, followed by activation of the cooling water for the superconducting section of the accelerator. 
\subsection{Design of stub shielding walls}
The general procedure for penetration calculations was as follows. First the PHITS 3.1 accelerator source term model was used to collect tracks at a selected surface, which were stored in an MCPL file. This information was placed in a ROOT multi-dimensional histogram and re-sampled as a new source term for MCNP 6.2. If weight-windows were further needed in the second calculation, they were generated using ADVANTG. \\
\indent Along the length of the accelerator there are 27 stubs which are used for the routing of electricity and RF wave-guides to the the tunnel along the accelerator length. The first few stubs have an opening of 150 cm in height and 180 cm in width while the remaining are 120 cm in width and with the same height. Fig. 2 shows a vertical cut through an example of one of the stubs. The four corners of the chicane also contain electrical filler material, which are not shown in the figure. The wave-guides themselves were also included in the model, but not shown in the figure. From a radiation protection point of view, the stubs are of particular interest because they are connected to the Klystron gallery, which is accessible during normal operations of the facility. This sets a dose rate requirement of 3 $\mu$Sv/hr in the Klystron gallery, which meant additional shielding would be needed. \\
\indent An example of the neutron energy spectrum entering the third leg of the chicane in the 2 GeV section of the accelerator is shown in Fig. 3. It can be seen that a sizable high-energy component penetrates the berm between the accelerator tunnel and the kylstron gallery. The neutron dose rates around this location are shown in Figure 4. It was found that a 50 cm thick, 200 cm high, 200 cm wide normal concrete wall plus 10 cm thickness of heavy concrete would be needed in order to bring the total dose rate below the limit of 3 $\mu$Sv/hr. A safety factor of 2 is further required for ESS shielding calculations, thus the design limit is 1.5 $\mu$Sv/hr. \\
\begin{figure}[t]
  \centering
  \resizebox{0.4\textwidth}{!}{\includegraphics{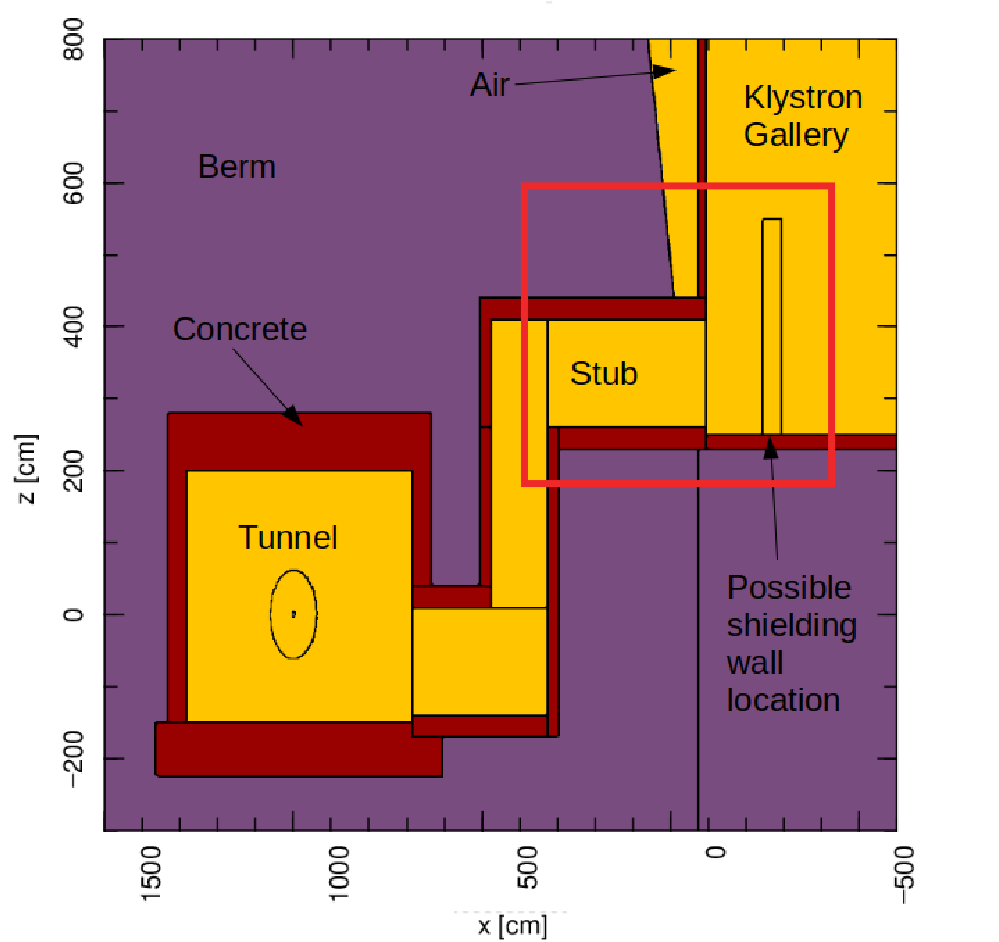}}
  \caption{\label{fig:3} A vertical cut through one of the stubs and indicating a potential location for the shielding wall. The wall placement indicates the approximate position and design before the optimization was carried out.}
\end{figure}
\begin{figure}[t]
  \centering
  \resizebox{0.45\textwidth}{!}{\includegraphics{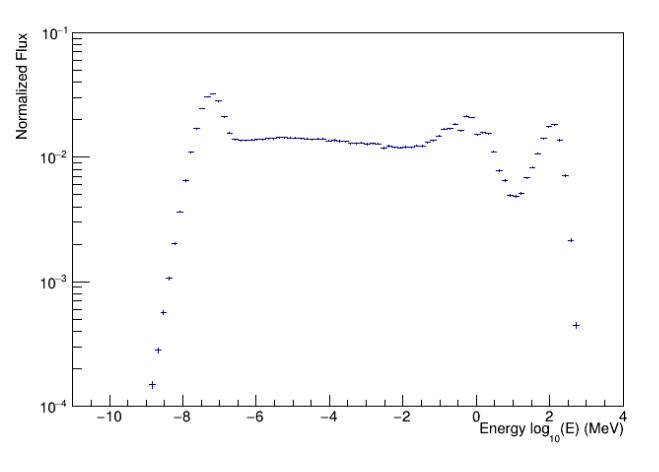}}
  \caption{\label{fig:3} The neutron energy spectrum entering a stub in the 2 GeV section of the accelerator.}
\end{figure}
\begin{figure}[t]
  \centering
  \resizebox{0.45\textwidth}{!}{\includegraphics{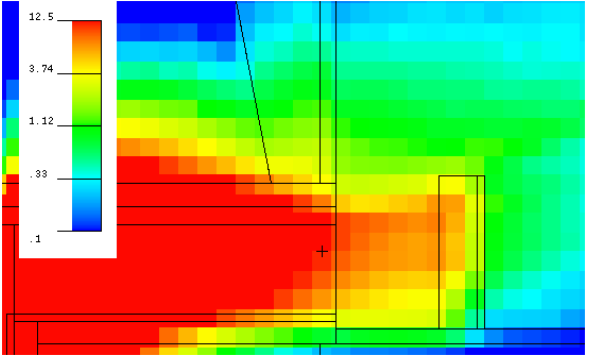}}
  \caption{\label{fig:3} The neutron dose rate around the third leg of a stub in the 2 GeV section of the accelerator. The plot is in the indicated red boxed region in Fig. 2. The Units are
  $\mu$Sv/hr.}
\end{figure}
\subsection{Cooling water activation}
Water for cooling of the accelerator components circulates from the water cooling room, adjacent to the kylstron gallery, into the accelerator tunnel and back.
During operation of the accelerator, workers have access to this room and thus dose rate calculations in the vicinity of the water loops is needed.
Such calculations are challenging due to the length of the accelerator, changing beam energy and the number of water volumes present.
This is further complicated by the fact that the activation products move with the water flow, which is not easily modeled satisfactorily in Monte-Carlo codes,
which can only describe stationary configurations. Of particular importance is the water in the magnets of the superconducting section of the accelerator,
which was found to be the dominant source in the water cooling room during normal operation, and thus is the focus in the following. \\
\indent The basic premise for the calculation of the dose rates in the water cooling room is described in \cite{masukawa2009}. In this work, representative
water volumes along the accelerator were defined in the CombLayer model. This included the return and supply pipes, and water in the magnets and the RF couplers.
The neutron spectra and isotope yields were calculated in these volumes and DCHAIN-SP was used to calculate the activation in the individual volumes.
The water however circulates and the following models were used, depending on the half-lives of the isotopes. For isotopes with half-lives comparable to (or larger) the operation time, $T_o$, of 100 days, the activation for the ith component of the loop can be calculated by
\begin{equation}
  A_i[Bq]=N_i{\sigma}\Phi_i[1-e^{-{\lambda}T_o}] ,
\end{equation}
where $N_i$ is the number of un-activated atoms in the ith component of the loop, $\sigma$ is the neutron cross-section for a given isotope, $\lambda$ is the decay constant for a given isotope, and $\Phi_i$ is the neutron flux in the ith component of the loop. This equation is directly solved by DCHAIN-SP by setting the irradiation time to 100 days. For the short-lived isotopes, the equilibrium activity in the water cooling room is calculated and given by
\begin{equation}
  A_i[Bq]=\frac{N_i{\sigma}\Phi_i[1-e^{-{\lambda}t_{ir,i}}]e^{t_{c,i}}}{1-e^{-\lambda T}} ,
\end{equation}
where $t_{ir,i}$ is the irradiation time in the component, $t_{c,i}$ is the cooling time from the end of the component to the water cooling room, and $T$ is the total circulation time around the loop. The total equilibrium activity was calculated from the sum of the individual components in the loop. In the calculations, the average circulation times and irradiation times were estimated from available engineering data. \\
\indent The calculation of the activities in this way represents the activation for a volume of water corresponding to the entire loop. To get the photon dose rates in the water cooling room, the total activity of the full loop volume was scaled to representative volumes of pipes in the water cooling room. A photon source term was then
created for MCNP6.2 based on the output data from DCHAIN-SP and used to estimate the dose rates on the surface of the pipes, which was found to be around
2-3 $\mu$Sv/hr per pipe. Thus in areas where pipes are grouped together, some countermeasures may be needed. The isotopes and their activities
contributing to these dose rates are shown in table 1. \\
\begin{table}[htb]
  \centering
  \caption{Activity of isotopes contributing to the dose rates in the water cooling room from cooling water of the superconducting accelerator
    components after 100 days of normal operations and zero days of cooling. Corrosion products are not included in the analysis.}
  \begin{tabular}{ll|ll}\toprule
  Isotope    & Activity (Bq/m$^3$) &    Isotope    & Activity (Bq/m$^3$)  
\\ \midrule
$^{7}$Be  & 2.7$\times10^{7}$ & $^{17}$N  & 3.9$\times10^{1}$
\\
$^{11}$Be & 1.5$\times10^{4}$ & $^{14}$O  & 4.4$\times10^{6}$
\\
$^{10}$C  & 2.5$\times10^{5}$ & $^{15}$O  & 1.3$\times10^{7}$
\\
$^{11}$C  & 4.5$\times10^{6}$ & $^{19}$O  & 7.8$\times10^{2}$
\\
$^{15}$C  & 4.6$\times10^{-1}$ & $^{17}$F  & 1.9$\times10^{4}$
\\
$^{13}$N  & 2.4$\times10^{6}$ & $^{18}$F  & 8.1$\times10^{3}$
\\
$^{16}$N  & 7.0$\times10^{4}$ & $^{19}$Ne  & 1.7$\times10^{2}$
\\
\bottomrule
\end{tabular}
  \label{tab:widetable}
\end{table}

\section{Conclusions}
In conclusion, extensive Monte-Carlo calculations have been carried out of the ESS accelerator environment.
These include calculation of neutrons and photon transport through various chicanes and activation of accelerator components.
For these calculations, new methods were developed in order to make use of extensive software developments in the field.
In addition to the work shown here, calculations have also been carried out for numerous chicanes around the accelerator,
including the alignment penetrations, the front-end building penetrations, and a gallery for access for the CTL.
The models have also been used for estimates of full beam losses around the facility, skyshine calculations, and in addition can be used in the future
for estimates of the free release of activated components.

\section{Acknowledgments}
The authors would like to thank Thomas Kittelmann for the support in developing
the MCPL PHITS interface.

\bibliographystyle{ans}
\bibliography{example}
\end{document}